\documentclass[aps,prl,twocolumn,showpacs,superscriptaddress,bibnotes]{revtex4}

\usepackage{graphicx}
\usepackage{dcolumn}
\usepackage{bm}

\begin{document}

\preprint{submitted to PRL}

\title{Three-Dimensional Bulk Electronic Structures of 
Ca$_{1.5}$Sr$_{0.5}$RuO$_{4}$ Studied 
by Soft X-ray Angle-Resolved Photoemission}

\author{M. Uruma}
\author{A. Sekiyama}
\author{H. Fujiwara}
\author{M. Yano}
\author{H. Fujita}
\author{S. Imada}
\affiliation{Division of Materials Physics, 
Graduate School of Engineering Science, Osaka 
University, Toyonaka, Osaka 560-8531, Japan}
\author{T. Muro}
\affiliation{Japan Synchrotron Radiation Research Institute, SPring-8, 
Mikazuki, Hyogo 679-5198, Japan}
\author{I. A. Nekrasov}
\affiliation{Institute of Electrophysics, Russian Academy of 
Sciences-Ural Division, 626041 Yekaterinburg, GSP-170, Russia}
\author{Y. Maeno}
\affiliation{Department of Physics, 
Kyoto University, Kyoto 606-8502, Japan}
\author{S. Suga}
\affiliation{Division of Materials Physics, 
Graduate School of Engineering Science, Osaka 
University, Toyonaka, Osaka 560-8531, Japan}

\date{\today}

\begin{abstract}
We report on experimental data
of the three-dimensional bulk Fermi surfaces 
of the layered strongly correlated 
Ca$_{1.5}$Sr$_{0.5}$RuO$_4$ system.
The measurements have been performed by means of 
$h\nu$-depndent bulk-sensitive soft x-ray angle-resolved photoemission 
technique. 
Our experimental data evinces the 
bulk Fermi surface topology at $k_z\sim$0 to be qualitatively different 
from the one observed by surface-sensitive low-energy ARPES. 
Furthermore, stronger $k_z$ dispersion of the circle-like $\gamma$ 
Fermi surface sheet is observed compared with Sr$_2$RuO$_4$.
Thus in the paramagnetic metal phase, 
Ca$_{1.5}$Sr$_{0.5}$RuO$_4$ compound is found to have 
rather three-dimensional electronic structure.

\end{abstract}


\maketitle

Strongly correlated transition metal oxides are widely studied 
because of a variety of such intriguing phenomena 
as (high-temperature) anisotropic superconductivity, 
Mott transition, magnetic and/or orbital ordering, and 
large mass enhancement. 
Among them, the single-layered perovskite Ca$_{2-x}$Sr$_x$RuO$_4$ is 
particularly interesting, since it shows various phases 
as functions of temperature and $x$~\cite{Maeno,Nakatsuji}. 
By substituting Sr$^{2+}$-ions for isovalent Ca$^{2+}$-ions, 
the unconventional superconductivity takes place 
below $\sim$1.5 K for $x=2$. 
A paramagnetically metallic behavior with strong electron 
correlation is seen in a wide temperature region for $0.5<x<2$. 
The system shows a paramagnetic metal to 
``magnetic'' metal transition at 10 K, for $0.2<x<0.5$, 
and eventually becomes a Mott insulator for $x<0.2$. 
At the Sr concentration of $x=0.5$, 
it is furthermore known that the effective mass diverges below 10 K 
although the system does not undergo 
a Mott transition~\cite{Nakatsuji,cluster}. 
These phenomena are thought to originate  
not only from the enhanced electron correlation 
by the rotation of the RuO$_6$ octahedra
(Ru-O-Ru bonds are bent from ideal ones), 
but also from the reduction of orbital degree of freedom 
by a Jahn-Teller distortion~\cite{OSMT, oo1, oo2, oo3} 
and/or a Ru 4$d$ spin-orbit interaction. 
Indeed, a (partial) Ru 4$d$ orbital-ordering has been reported 
(proposed) 
for $x = 0$~\cite{Zegkinoglou} ($0.2<x<0.5$~\cite{cluster, OSMT}). 

Angle-resolved photoemission (ARPES) is a very powerful tool 
to evince band dispersions and  Fermi surface (FS) topology. 
So far, several low-$h\nu$ ARPES experiments were performed 
for Ca$_{2-x}$Sr$_x$RuO$_4$~\cite{low1,KMShen,surface,low2}. 
However, it is known that low-energy 
photoemission is surface-sensitive and 
often provides spectral shapes which are not consistent 
with bulk electronic structures
~\cite{HE1,HE2,PRL93,Suga1}. 
Meanwhile, it has been demonstrated that 
high-$h\nu$ ARPES with use 
of the soft x-ray can reveal detailed band dispersions of 
the bulk electronic states~\cite{SekiyamaPRB,Suga2,Yano}. 
In addition, by virtue of the longer mean free path 
of higher kinetic energy photoelectrons 
the high-energy $h\nu$-dependent ARPES 
can probe electronic structures of three-dimensional compounds
with better-resolved $k_z$ dispersion ($k_z$ - momentum component 
perpendicular to the cleaved sample surface)~\cite{Yano,MFP,LSFO}. 
%
%
%

High quality single crystal of 
Ca$_{1.5}$Sr$_{0.5}$RuO$_{4}$ ($x$ = 0.5), 
which was grown by the floating zone method~\cite{FZM}, 
was used for the measurement. 
All ARPES measurements were performed at BL25SU 
in SPring-8~\cite{Saitoh}. 
The high-energy ($h\nu$ = 708 eV) and low-energy ($h\nu$ = 362 eV) 
soft x-ray ARPES and FS mapping were carried out 
for the $k_z\sim$0 plane. 
In addition, $h\nu$ dependent ($h\nu$ = 650-730 eV) ARPES 
for the $k_x$=0 plane FS mapping 
was done with the energy step of 5 eV. 
The base pressure was about $4\times 10^{-8}$ Pa. 
The (001) clean surface was obtained by cleaving 
the sample \textit{in situ} 
at the measuring temperature of 20 K. 
The overall energy resolution was set to 200 meV. 
The angular resolution was $\pm$0.15$^{\circ}$($\pm$0.25$^{\circ}$) 
for the parallel (perpendicular) direction to the analyzer slit. 

%
%
%
\begin{figure}
\includegraphics[width=7.5cm,clip]{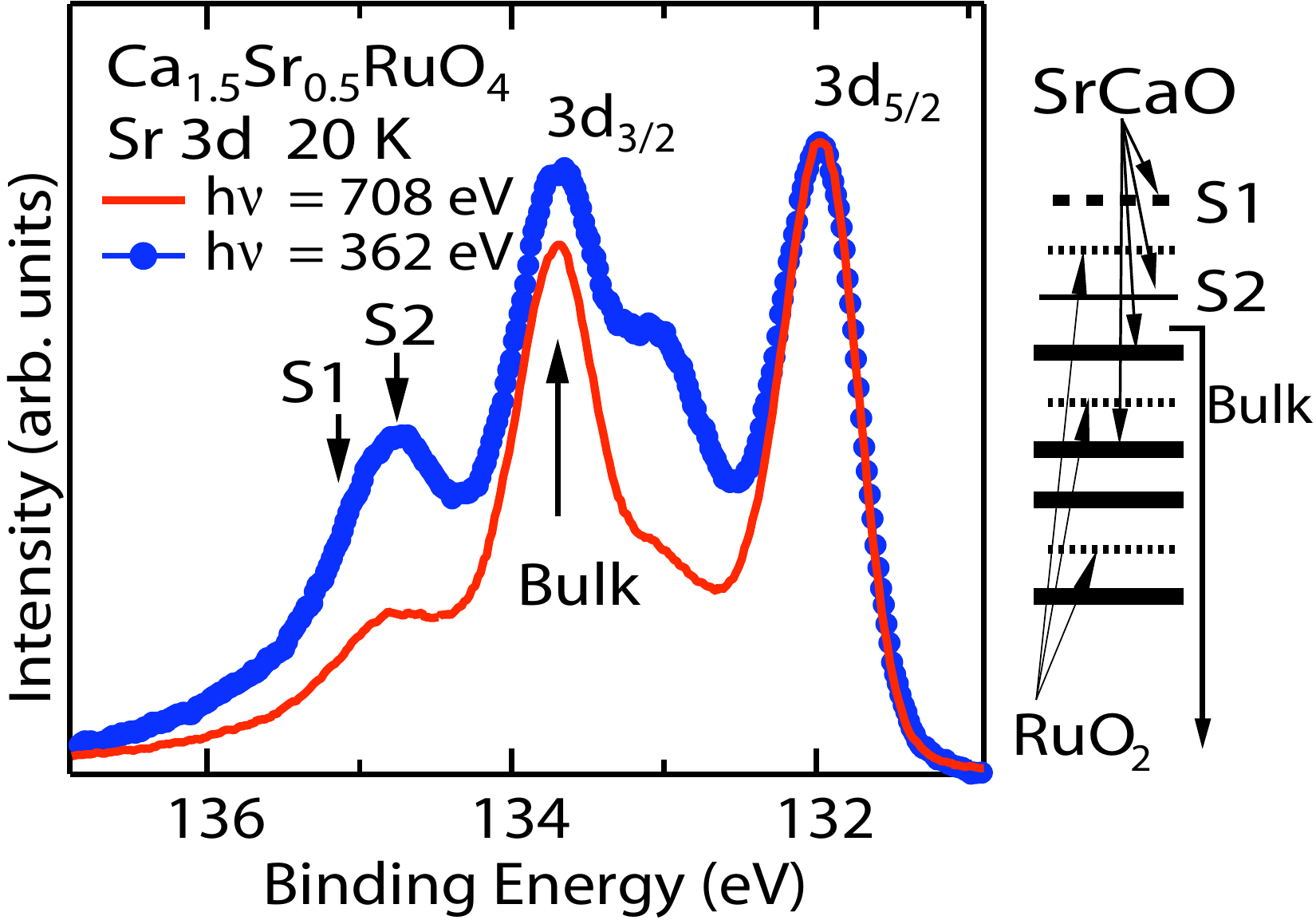} %
\caption{\label{Fig1} 
(color online) Left panel: $h\nu$ dependence of the Sr 3\textit{d} 
core-level PES spectra of 
Ca$_{1.5}$Sr$_{0.5}$RuO$_4$ with the energy resolution of 200 meV. 
The spectra are normalized at the $3d_{5/2}$ main peak. 
The dots with line and solid line correspond to $h\nu$= 362 eV and 
708 eV.
Right panel: schematic picture
of atomic layeres. Strong surface contribution is 
expected from top-most SrCaO surface layer (S1) and 
the second SrCaO surface layer (S2) 
placed just below the top RuO$_2$ layer.
}

\end{figure}

Figure 1 shows the $h\nu$ dependence of the angle-integrated Sr 3$d$ core-level 
photoemission (PES) spectra of Ca$_{1.5}$Sr$_{0.5}$RuO$_4$. 
To our understanding these spectra contain three components corresponding to 
the contributions from the top-most SrCaO surface layer (S1), 
the second SrCaO surface layer (S2) located 
just below the top RuO$_2$ layer, 
and the bulk layers~\cite{SekiyamaPRB}. 
The intensity of the S1 and S2 shoulders 
is observed to be remarkably stronger
for $h\nu$ = 362 eV (blue curve) than for $h\nu$ = 708 eV (red curve). 
To estimate the bulk contribution to the spectra
we did line-shape analysis in the same manner as in 
Ref.~\cite{SekiyamaPRB}. 
The bulk contribution is estimated to be 55\% (39\%) 
in the Sr $3d$ spectra at $h\nu$ = 708 (362) eV.
These values are close to the ones evaluated 
for the given lattice constants 
and the calculated photoelectron mean free path $\lambda$~\cite{MFP}. 
Then, the bulk contribution in the valence-band is found to be
$\sim$65 \% at $h\nu$ = 708 eV and $\sim$50 \% 
at $h\nu$ = 362 eV, manifesting that the 
708 eV-PES mainly probes the bulk valence states.

\begin{figure}
\includegraphics[width=8cm,clip]{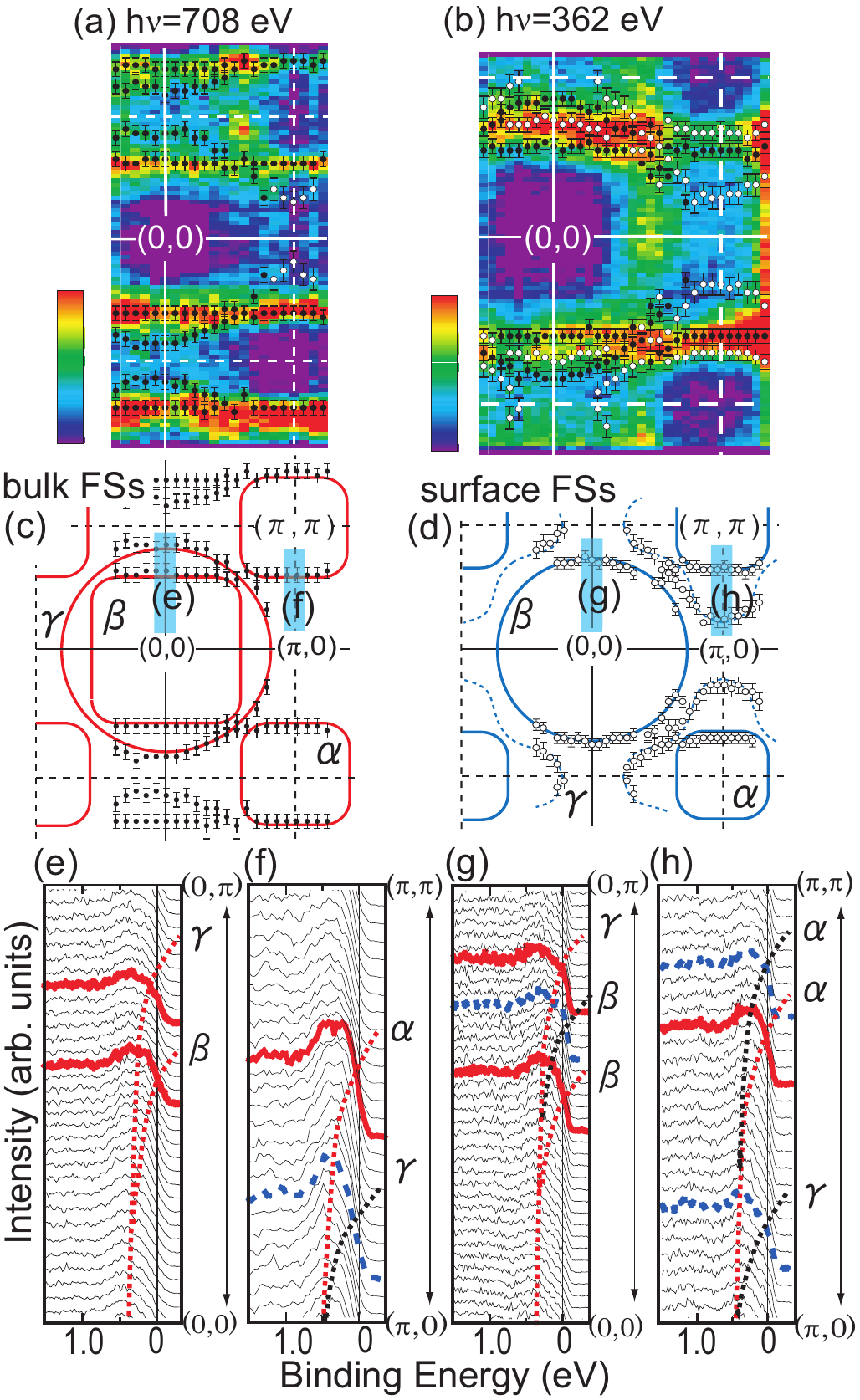}
\caption{\label{Fig2}(color online) 
(a),(b) Experimental Ca$_{1.5}$Sr$_{0.5}$RuO$_4$ FS 
intensity map at $k_z\sim$0 obtained by the 
708 eV-ARPES and 362 eV-ARPES, respectively. 
The filled and empty circles represent the experimental
$k_F$ values for the bulk and surface as discussed in the text. 
(c),(d) Schematic FS (red, blue lines) with respect to (a),(b). 
The surface FS observed in (a) is not drawn in (c) and 
the bulk FSs revealed in (b) are not shown in (d) for simplicity. 
The blue dashed line in (d) reproduces the low-energy ARPES data 
of Ref.~\cite{surface}. 
(e),(f): 708 eV-ARPES spectra along (0,0)-(0,$\pi$) and 
($\pi$,0)-($\pi$,$\pi$) directions within shaded areas in (c),(d). 
(g),(h): 362 eV-ARPES data along the same direction as (e),(f). 
Red dashed and black dashed lines are experimental band 
dispersions for bulk and surface, which are evaluated by the 
momentum distribution curves (MDCs). 
The red-bold and blue-dashed lines in (e)-(h) correspond to 
bulk and surface spectra at $k_F$.
The BZ corresponds to ideal two-dimensional cubic lattice in which 
the effect of RuO$_6$ rotation is not taken into account. 
}
\end{figure}

In the following, we show the FSs  derived 
from the 708 eV- and 362 eV-ARPES spectral intensity integrated 
from $-$0.1 to 0.1 eV with respect to the Fermi level ($E_F$) 
(Figs.~\ref{Fig2} (a) and (b)). 
These results are obtained from energy distribution curves (EDCs) 
(Figs.~\ref{Fig2} (e)-(h)) 
in the paramagnetic metal phase at 20 K. 
Figures 2(a) and 2(b) display raw data 
and Figs. 2(c) and 2(d) summarize remarkable 
features (see captions). 
We have obtained $k_F$ 
($E_F$-crossing points in a reciprocal space) 
from both EDCs and momentum distribution curves 
(MDCs: not shown here) 
as plotted in Figs.~\ref{Fig2} (a) and (b) 
with a two-dimensional cubic symmetry BZ. 
Although the true BZ of Ca$_{1.5}$Sr$_{0.5}$RuO$_4$ is 
$\sqrt2$$\times$$\sqrt2$ folded 
because of the rotation RuO$_6$ octahedra along the $c$-axis 
($\sim$12$^\circ$~\cite{oo1}), 
we use the size of BZ and notations for high symmetry directions 
of the Sr$_{2}$RuO$_4$ crystal structure. 
EDCs in Figs.~2(e)-(h) correspond to shaded areas in 
Figs.~2(c) and (d). 
As shown in Fig.~2(e), two branches ($\beta$ and $\gamma$) 
located at 0.5 eV at (0,0) 
approach and cross $E_F$ between (0,0) and (0,$\pi$) 
in the 708 eV-ARPES spectra. 
In the 362 eV-ARPES spectra in Fig.~2(g), 
we observed one additional branch between these two 
branches crossing $E_F$. 
Considering that the 362 eV-ARPES is more surface sensitive, 
we attribute this additional branch 
(which spectral weight 
is noticeably suppressed in the 708 eV-ARPES spectra) 
to be resulting from  the surface band 
different from the bulk one.
Along the ($\pi$,0)-($\pi$,$\pi$) direction, 
we have likewise found that two 
branches cross $E_F$ in the 708 eV-ARPES spectra (Fig. 2(f)), 
whereas three 
branches cross $E_F$ in the 362 eV-ARPES spectra (Fig. 2(h)). 
Hence essentially dissimilar electronic structures
and Fermi surface topology (at $k_z\sim$0)
between the bulk and surface states
for this ``quasi'' two-dimensional electron system
are demonstrated.

As shown in Figs. 2(a) and 2(c) 
by 708 eV-ARPES measurements, 
we have found three bulk FS sheets for 
Ca$_{1.5}$Sr$_{0.5}$RuO$_4$, namely, 
one hole-like square-shaped $\alpha$ sheet centered 
at ($\pi$,$\pi$), 
one electron-like square-shaped $\beta$ sheet 
and one electron-like circle-shaped $\gamma$ sheet 
both centered at (0,0). 
Our FS topology observed for $k_z\sim$0 plane 
is thus very similar to that of 
bulk sensetive experimental data for Sr$_2$RuO$_4$ and 
Ca$_{0.2}$Sr$_{1.8}$RuO$_4$~\cite{SekiyamaPRB}. 
If one assume that orbital character for the FS sheets remains 
the same as for Sr$_2$RuO$_4$, 
then $\alpha$ and $\beta$ sheets should correspond mainly to 
the Ru $4d_{yz/zx}$ orbitals whereas the $\gamma$ sheet 
has mainly the Ru $4d_{xy}$ character. 
Possible effects of the BZ folding caused 
by the RuO$_6$ rotation~\cite{EKo} 
are not seen (or negligibly weak) in our results 
in strong contrast to the low-energy ARPES result 
for Sr$_2$RhO$_4$ and Sr$_2$RuO$_4$ surface~\cite{SRhO,KMShen}. 
The reason for the absence of the BZ folding effect 
is, however, not clear at present. 

On the other hand, the 362 eV-ARPES results (Fig. 2(b)) 
clearly reveal three surface FS sheets (open circles) distinguished 
from the bulk FS sheets. 
While the shape of the surface $\alpha$ sheet 
is rather similar to the bulk one as recognized by comparing Figs.~2(c) and 
(d), the electron-like surface $\beta$ sheet 
is much more circle-like compared to 
the bulk $\beta$ sheet. 
Furthermore, the FS topology of the $\gamma$ sheet is 
qualitatively different between the bulk in Fig.~2(c) and surface in Fig.~2(d). 
The hole-like surface $\gamma$ sheet 
observed here well traces the result of the low-energy ARPES 
at $h\nu$ = 32 eV~\cite{surface} reproduced by the dashed curves in Fig.~2(d). 
The scanning tunneling microscopy (STM) measurement of Sr$_2$RuO$_4$ 
has shown a surface reconstruction~\cite{Matzdorf}. 
Except for this report no surface reconstruction has been reported 
for Ca$_{2-x}$Sr$_x$RuO$_4$. 
It has been believed that the bulk FSs can be detected 
even by the surface-sensitive 
low-energy ARPES for Ca$_{2-x}$Sr$_x$RuO$_4$ since 
the RuO$_6$ rotation angle indicates similar
crystal structures between 
the bulk ($\sim$12$^\circ$) and surface ($\sim$11$^\circ$)
~\cite{surface}. 
However, our result evinces that
the surface FS topology is noticeably different from the bulk. 
%
%
\begin{figure}
\includegraphics[width=8.5cm,clip]{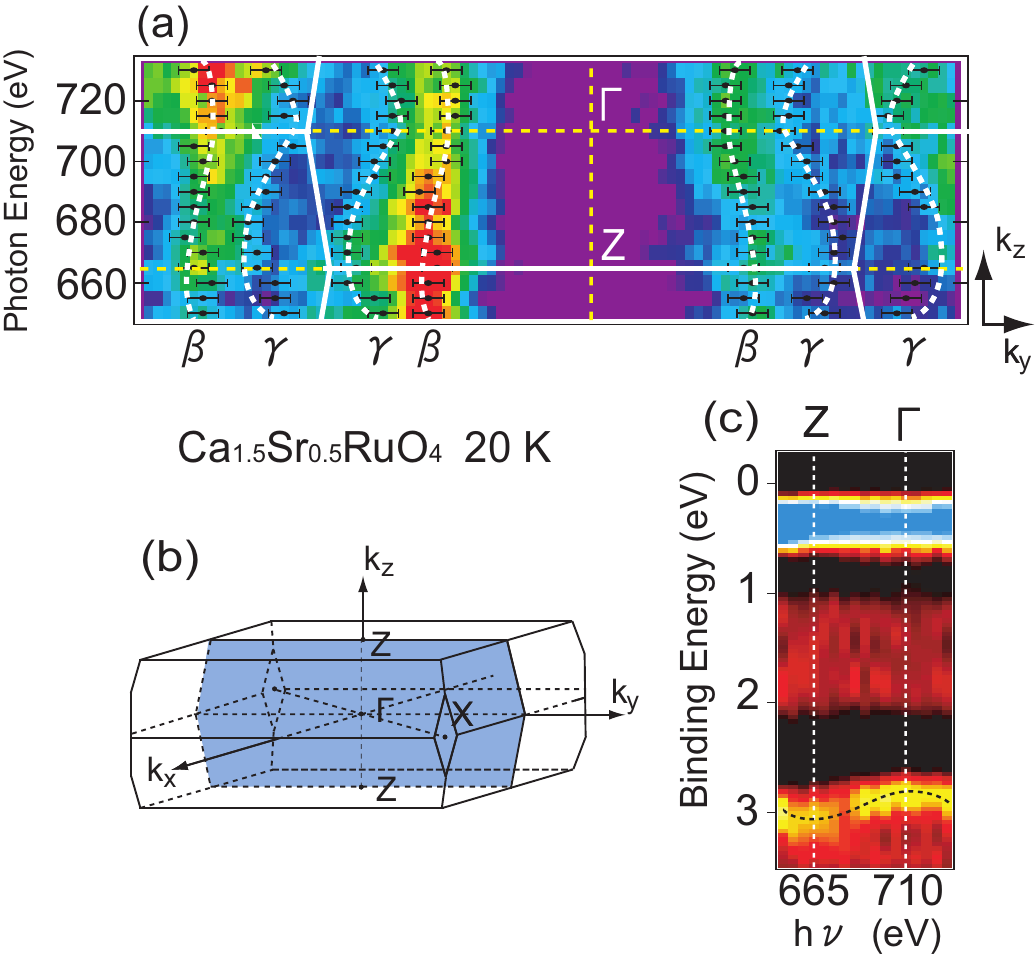}
\caption{\label{Fig3}(color online) 
(a) Obtained by $h\nu$-dependent ARPES 
measurements intensity map integrated over $E_F \pm 0.1$ eV 
for Ca$_{1.5}$Sr$_{0.5}$RuO$_{4}$ in the $k_x$=0 plane. 
The black dots with the error bars represent 
experimental $k_F$ values. 
The white solid lines represent part of tetragonal BZ boundaries 
shown in (b). 
(c) Second derivative of the EDCs along the $\Gamma$-Z 
direction. The black dashed curve near 3 eV is the guide to the eye.
}
\end{figure}

Below we discuss experimental data on interlayer coherent
electron hopping or in other words $k_z$ dispersion
of the FS observed by our bulk sensetive $h\nu$ dependent ARPES measurements.
A nearly free photoelectron model calculation 
with the inner potential of 9 eV identifies
$\Gamma$ and Z points of a tetragonal BZ 
with $h\nu$s of 712 and 664 eV in our experiment~\cite{comment}. 

Figure~\ref{Fig3}(a) shows the results of $h\nu$-dependent 
ARPES for Ca$_{1.5}$Sr$_{0.5}$RuO$_{4}$, 
representing the $k_z$ dispesion of 
the $\beta$ and the $\gamma$ FS sheets. 
The $\gamma$ FS sheet is found to have significant $k_z$ dependence 
reflecting stronger three-dimensionality 
of the Ca$_{1.5}$Sr$_{0.5}$RuO$_{4}$ electronic structure. 
This is in strong contrast 
to results of quantum oscillation measurements 
for Sr$_2$RuO$_4$ showing negligible 
$k_z$ dependence for the $\gamma$ sheet~\cite{dHvA,Mackenzie}. 
On the other hand, our result has revealed
much weaker $k_z$ dependence of the $\beta$ FS sheet compared 
with that of the $\gamma$ FS sheet. 
Indeed, a ratio of $k_F$ between the minimum and maximum 
along the $\Gamma$-Z direction
has been estimated 
to be $\sim$0.9 for the $\beta$ 
sheet whereas for the $\gamma$ sheet
it is $\sim$0.8. 
So far it has  been considered that the electronic structures and 
FSs of such single-layered transition metal oxides as 
Ca$_{2-x}$Sr$_x$RuO$_4$ are highly two-dimensional. 
However, those for Ca$_{1.5}$Sr$_{0.5}$RuO$_4$ are found to 
be more three-dimensional than Sr$_{2}$RuO$_{4}$ from our experiments. 

As shown in Fig. 3(c), 
the spectral weight near 3 eV in the ARPES data 
disperses clearly as a function of $h\nu$.
Its bottom (top) is located at Z ($\Gamma$). 
This dispersion is qualitatively consistent with results of 
the band-structure calculation for Sr$_2$RuO$_4$ 
along the $\Gamma$-Z ($k_z$) direction~\cite{Oguchi,dHvA,SRODMFT}. 
However there seems to be also non-dispersive contribution at 3 eV.
These $k$-independent spectral structure observed 
at $\sim$1-2 and 3 eV can be presumably identified as the  
part of the Ru 4d(t$_{2g}$) lower Hubbard band similar to our recent
LDA+DMFT(QMC) results for Sr$_2$RuO$_4$~\cite{SRODMFT}.
Namely, correlations in Ca$_{1.5}$Sr$_{0.5}$RuO$_4$
can be enhanced due to distortion in comaprison with Sr$_2$RuO$_4$.

Since the bulk FS topology of Ca$_{1.5}$Sr$_{0.5}$RuO$_4$
at $k_z\sim$0 observed by us is similar to Sr$_2$RuO$_4$
one can expect the orbital characters of the relatively 
large $\gamma$ and small $\beta$ sheets to be of the Ru $4d_{xy}$ 
and $4d_{yz/zx}$ symmetries, respectively. 
In that way our results seem to contradict with theoretical 
expectation 
where strong $k_z$ dependence is usually expected for 
the Ru $4d_{yz/zx}$-derived bands rather than for the Ru $4d_{xy}$ band~\cite{Oguchi,dHvA,SRODMFT}. 
One possible explanation of our ARPES results is that 
the characters of the $\beta$ and $\gamma$ FS sheets are interchanged with 
each other compared with those for Sr$_2$RuO$_4$. 
Our LDA (local density approximation) band-structure calculation~\cite{LDA} 
as well as that in Ref.~\cite{EKo} predict such a scenario 
as the smallest electron-like two-dimensional FS sheet originats 
from the Ru $4d_{xy}$ band.
But this FS sheet is caused by doubling of the lattice 
in Ca$_{1.5}$Sr$_{0.5}$RuO$_4$. 
As mentioned above in our ARPES results, 
bands coming from folding are negligibly weak. 
We have surveyed not only the reciprocal space shown 
in Figs. 2 and 3 
but also the $k_x-k_y$ plane at $h\nu$ = 650 eV (not shown here) covering first and 
second BZs by ARPES, 
therefore the absence of 
these hole-like FS sheets is thought to be intrinsic 
in this experiment for Ca$_{1.5}$Sr$_{0.5}$RuO$_4$ system.
 
Anoher scenario of rather strong $k_z$ dependence of the $\gamma$ 
sheet can be by the effect of 
the RuO$_6$ rotation within the conducting plane. 
By virtue of the rotation, there appears small but finite 
$\sigma$-bonding between the Ru $4d_{xy}$ and the O $2p_{x/y}$ 
orbitals. 
So the situation becomes similar to high-T$_C$ cuprate La$_{2}$CuO$_4$~\cite{Pavarini}, 
where it was shown that such $\sigma$-bonding leads to essentially 
three-dimensional $x^2-y^2$-symmetry Wannier function. 
For Ca$_{1.5}$Sr$_{0.5}$RuO$_4$
this effect is weaker than for La$_{2}$CuO$_4$.
Nevertheless Ru $4d_{xy}$-derived band 
in the distorted crystal structure 
can form rather three-dimensional $\gamma$ sheet.
Also one should mention here that more two-dimensional square-shaped $\beta$ sheet 
has stronger nesting instability as observed in Sr$_2$RuO$_4$ and 
Ca$_{0.2}$Sr$_{1.8}$RuO$_4$~\cite{SekiyamaPRB,Sidis}. 
Thus the Ru $4d_{yz/zx}$ orbital ordering scenario 
within the magnetic metal phase for $0.2 < x < 0.5$~\cite{OSMT} is possible.

To summarize, we have observed three-dimensional Fermi surfaces 
of Ca$_{1.5}$Sr$_{0.5}$RuO$_{4}$ in the paramagnetic metal phase 
at 20 K by applying 
bulk sensetive soft x-ray $h\nu$-dependent ARPES technique. 
We have revealed the genuine bulk electronic structures 
which are different from surface sensetive data for this material.
FS topology in the $k_z\sim$0 plane 
is observed to be qualitatively similar with bulk FS of Sr$_2$RuO$_4$.
However remarkable $k_z$ dependence of 
the $\gamma$ FS sheet have proved 
Ca$_{1.5}$Sr$_{0.5}$RuO$_{4}$ to have more three-dimensional 
electronic structure than Sr$_2$RuO$_4$.

We are grateful to H. Higashimichi, 
G. Funabashi, and T. Nakamura for supporting the 
experiments. This work was supported 
by a Grant-in-Aid for Scientific 
Research (15GS0213, 1814007, 18684015) 
and the 21st COE program (G18) of 
the Japan Society for the Promotion of Science and MEXT. 
This work was also supported by 
Hyogo Science and Technology Association. 
The ARPES was performed at SPring-8 
under the approval of JASRI (2006A1169, 2007A1005). 
IN thanks RFFI grants 08-02-00021, 08-02-00712, 08-02-91200, 06-02-90537,
travel grant of UB RAS and grant of President of Russia MK-2242.2007.2.

\end{document}